\documentclass[aip,amsmath,amssymb,reprint]{revtex4-1}
\usepackage{graphicx}% Include figure files
\usepackage{dcolumn}% Align table columns on decimal point
\usepackage{bm}% bold math
\usepackage[utf8]{inputenc}
\usepackage[T1]{fontenc}
\usepackage{mathptmx}
\usepackage{etoolbox}

\makeatletter
\def\@email#1#2{%
 \endgroup
 \patchcmd{\titleblock@produce}
  {\frontmatter@RRAPformat}
  {\frontmatter@RRAPformat{\produce@RRAP{*#1\href{mailto:#2}{#2}}}\frontmatter@RRAPformat}
  {}{}
}%
\makeatother
\begin{document}

\preprint{AIP/123-QED}

\title[Unraveling nano-scale effects of topotactic reduction in LaNiO$_2$ crystals]{Unraveling nano-scale effects of topotactic reduction in LaNiO$_2$ crystals}

\author{Yu-Mi Wu}
 \altaffiliation{yu-mi.wu@cornell.edu}
 \altaffiliation[Current address: ]{Department of Materials Science and Engineering, Cornell University, Ithaca, New York 14853, USA}
\author{Pascal Puphal}
\author{Masahiko Isobe}
\author{Bernhard Keimer}
\author{Matthias Hepting}
\author{Y. Eren Suyolcu}
 \altaffiliation{eren.suyolcu@fkf.mpg.de}
\author{Peter A. van Aken}

\email{yu-mi.wu@cornell.edu and eren.suyolcu@fkf.mpg.de}

\affiliation{Max Planck Institute for Solid State Research, Heisenbergstrasse 1, 70569 Stuttgart, Germany}

\date{\today}

\begin{abstract}
Infinite-layer nickelates stand as a promising frontier in the exploration of unconventional superconductivity. 
Their synthesis through topotactic oxygen reduction from the parent perovskite phase remains a complex and elusive process. 
This study delves into the nano-scale effects of the topotactic lattice transformation within LaNiO$_2$ crystals. 
Leveraging high-resolution scanning transmission electron microscopy and spectroscopy, our investigations uncover a panorama of structural alterations, including grain boundaries and coherent twin boundaries, triggered by reduction-induced transformations.
In addition, our analyses unveil the formation of an oxygen-rich disordered transition phase encircling impurities and pervading crystalline domains and the internal strain is accommodated by grain boundary formation.
By unraveling these nano-scale effects, our findings provide insights into the microscopic intricacies of the topotactic reduction process elucidating the transition from the perovskite to the infinite-layer phase within nickelate bulk crystals. 
%This study sheds light on the elusive pathways of structural evolution crucial for advancing our understanding of these intriguing materials.
\end{abstract}

\maketitle

\section{Introduction}

The discovery of superconductivity in thin films of the infinite-layer (IL) nickelate Nd$_{0.8}$Sr$_{0.2}$NiO$_2$ \cite{li2019superconductivity} has opened a new avenue to explore a material system closely related to cuprate high-temperature superconductors \cite{Keimer2015}. 
The structure of IL nickelates contains quasi-two-dimensional NiO$_2$ planes, obtainable through topotactic oxygen reduction of the perovskite phase using either H$_2$ gas or metal hydrides as reducing agents \cite{li2019superconductivity,zeng2020phase,GaoCPL2021,Hayward1999,Crespin2005}. As an alternative route for oxygen deintercalation, the deposition of an aluminum capping layer on nickelate thin films has been employed recently \cite{Wei2023PRM,Wei2023SciAdv}.  
The nominal electronic configuration ($3d^9$) of the monovalent Ni$^{1+}$ cations in the NiO$_2$ planes is isolelectronic to that of Cu$^{2+}$ ions in undoped cuprates. However, the electronic structure of the material appears to be different from that of cuprates \cite{Hepting2021,goodge2021doping,rossi2021orbital,botana2020similarities,been2021electronic,Held2022,Chen2022}. 

The manifestation of superconductivity in IL nickelate thin films has exhibited consistent reproducibility, extending recently across an entire family of nickelates encompassing different rare-earth ions, $R$ = La and Pr \cite{Osada2020,Osada2021,Wang2022}, diverse levels of Sr and Ca-substitution levels \cite{Li20201,zeng2021}, and a range of substrates \cite{ren2022,lee2023linear}. Intriguingly, superconductivity has also been observed in the quintuple-layer nickelate Nd$_6$Ni$_5$O$_{12}$ without Sr or Ca substitution, which can be considered as a self-doped version of IL nickelates \cite{pan2022}. Additionally, La$_3$Ni$_2$O$_7$ was found to become superconducting under high pressure \cite{sun2023}, although its $3d^{7.5}$ electronic configuration deviates significantly from that of cuprates. This intricate panorama unveils tantalizing commonalities and discrepancies between nickelates and cuprates, underscoring the need for a comprehensive exploration to elucidate the full landscape of their behaviors.

In experimental practice, while the synthesis and fabrication of IL nickelates present significant challenges, these processes have been extensively improved for thin film samples \cite{lee2023linear,lee2020aspects,SunAdvMat2023,ChowFront2022}. Notably, during the topotactic reduction of thin films, the underlying substrate provides epitaxial support essential for fostering the IL phase formation. The thin film geometry, crucially, tends to yield a single orientation of the reduced phase and minimizes the emergence of impurity phases.
Nevertheless, a considerable expansion of the in-plane lattice constants and contraction of the out-of-plane lattice during the reduction poses a challenge for epitaxial films. This phenomenon often prompts the occurrence of extended defects as a means of alleviating epitaxial strain. 
In particular, strain-relieving Ruddlesden-Popper type stacking faults often occur densely in films grown on widely used SrTiO$_3$ substrates \cite{Li20201,Osada2021,zeng2021,lee2020aspects,Parzyck2024}. Other phases such as LaNiO$_{2.5}$ and $a$-axis oriented LaNiO$_2$ can also form in response to reduction-induced strain \cite{kawai2010,Osada2021}.
In general, the crystalline quality of the IL structure is found to be closely linked to that of the perovskite precursor.
Furthermore, for uncapped films, a decomposition of the infinite-layer phase in the uppermost regions of the films has been observed \cite{lee2020aspects,onozuka2016,ikeda2013}, preventing surface-sensitive techniques from the exploration of the structure-property relationships within these materials.

As part of alternative approaches, recent efforts have directed attention toward the synthesis of polycrystalline powders \cite{Li2020powder,Wang2020powder,Ortiz2022,Puphal2022Front} and IL nickelate single crystals \cite{puphal2021,puphal2023}.
These single crystals could, in principle, provide large single domains and cleavable surfaces.
However, in contrast to thin films benefiting from epitaxial support, the arrangement of NiO$_2$ planes within a crystal might align along any of three equivalent pseudocubic [001] directions of the perovskite phase. Consistent with this notion, three twin domains of the tetragonal $P4/mmm$ crystal structure were observed in topotactic La$_{1-x}$Ca$_x$NiO$_2$ crystals \cite{puphal2021}. Recently, the synthesis of topotactic LaNiO$_2$ crystals exceeding one millimeter in size was achieved, utilizing a direct contact method with the reducing agent CaH$_2$ \cite{puphal2023}. This method was subsequently optimized, employing cube-shaped LaNiO$_3$ single crystals, each with 1 mm edges, facilitating reproducible adjustment of the reduction parameters  \cite{hayshida2023}. Despite electron backscatter diffraction (EBSD) characterization of the resulting LaNiO$_2$ crystals has elucidated the distribution of 1-50 $\mu$m sized $P4/mmm$ twins across the crystal surface \cite{hayshida2023}, crucial insights into the topotactically transformed microstructure in the bulk of these crystals remain elusive. For a comprehensive understanding of the physical properties of topotactically transformed systems, it is critical to gain insights into the micro- and nanoscale impacts of the transformation process on the crystal lattice, as the invasive nature of the process can result not only in incompletely reduced regions within a crystal \cite{puphal2021}, but also in the emergence of unforeseen defects or phase decompositions \cite{hu2024atomic}. 

In the quest to unravel the complexities of the reduction process, spatially resolved scanning tunneling electron microscopy (STEM) combined with electron energy-loss spectroscopy (EELS) stands as a powerful tool to probe the local lattice structure, elemental composition, and electronic structure down to the atomic level, allowing for the identification and disentanglement of various phases in topotactic systems \cite{wu2023}. 
Here, we utilize the capabilities of high-resolution STEM-EELS to scrutinize the structural effects of the topotactic reduction in LaNiO$_2$ crystals. Nanoscopic and microscopic signatures of structural disorder and impurity phases throughout the reduced crystal are unveiled by atomic-resolution STEM imaging.
Our observations demonstrate coherent twin boundaries with a relative rotation of $\sim$80$^\circ$ as a consequence of strain accommodation from the pseudocubic to tetragonal phase transformation.
In addition, grain boundaries (GBs) formed by ultra-thin La$_2$O$_{3-\delta}$ slabs and Ni layers in parallel to the main crystallographic directions emerge within the crystal, likely originating from excessive reduction.  
At a macroscopic scale, an intricate network of cracks and GB-like regions separate single crystalline domains, while a structurally disordered shell encircles the impurity phase, \textit{i.e.} Ni inclusions.
Our local spectroscopic analyses reveal the oxygen-rich nature of these regions, highlighting the formation of a LaNiO$_{2+\delta}$ transition phase.
Together, our integrated structural and elemental analyses help to unveil the origin of these defects within topotactically reduced crystals, illuminating the intricate interplay between structure and strain in these systems.

\section{Methods}

%Crystal growth and topotactic reduction
Single crystals of perovskite LaNiO$_{3}$ were synthesized in a high-pressure optical floating zone (OFZ) furnace (model HKZ, SciDre GmbH, Dresden, Germany), as described in Refs. \cite{puphal2023,puphal2023phase}. The obtained centimeter-sized crystals were cut into smaller cube-like shaped crystals, with each surface of a cube corresponding to a pseudocubic [001] plane of the rhombohedral $R\bar3c$ crystal structure of LaNiO$_{3}$. The resulting cubes with dimensions of $\sim$1 mm$^3$ were subjected to a direct contact topotactic reduction with CaH$_2$, as described in Ref. \cite{hayshida2023}.

%STEM-EELS
Electron-transparent TEM specimens of the sample were prepared on a Thermo Fisher Scientific focused ion beam (FIB) using the standard liftout method. Samples with a size of 20 $\times$ 5 $\mu$m$^2$ were thinned to 30 nm with 2 kV Ga ions, followed by a final polish at 1 kV to reduce the effects of surface damage.
HAADF, ABF and EELS were recorded by a probe aberration-corrected JEOL JEM-ARM200F scanning transmission electron microscope equipped with a cold-field emission electron source and a probe Cs corrector (DCOR, CEOS GmbH), and a Gatan K2 direct electron detector was used at 200 kV. STEM imaging and EELS analyses were performed at probe semiconvergence angles of 20 and 28 mrad, resulting in probe sizes of 0.8 and 1.0 Å, respectively. Collection angles for STEM-HAADF and ABF images were 75 to 310 and 11 to 23 mrad, respectively. To improve the signal-to-noise ratio of the STEM-HAADF and ABF data while minimizing sample damage, a high-speed time series was recorded (2 $\mu$s per pixel) and was then aligned and summed. 

\section{Results}

%Figure 1
\begin{figure*}[t]
\centering
\includegraphics[width=\textwidth]{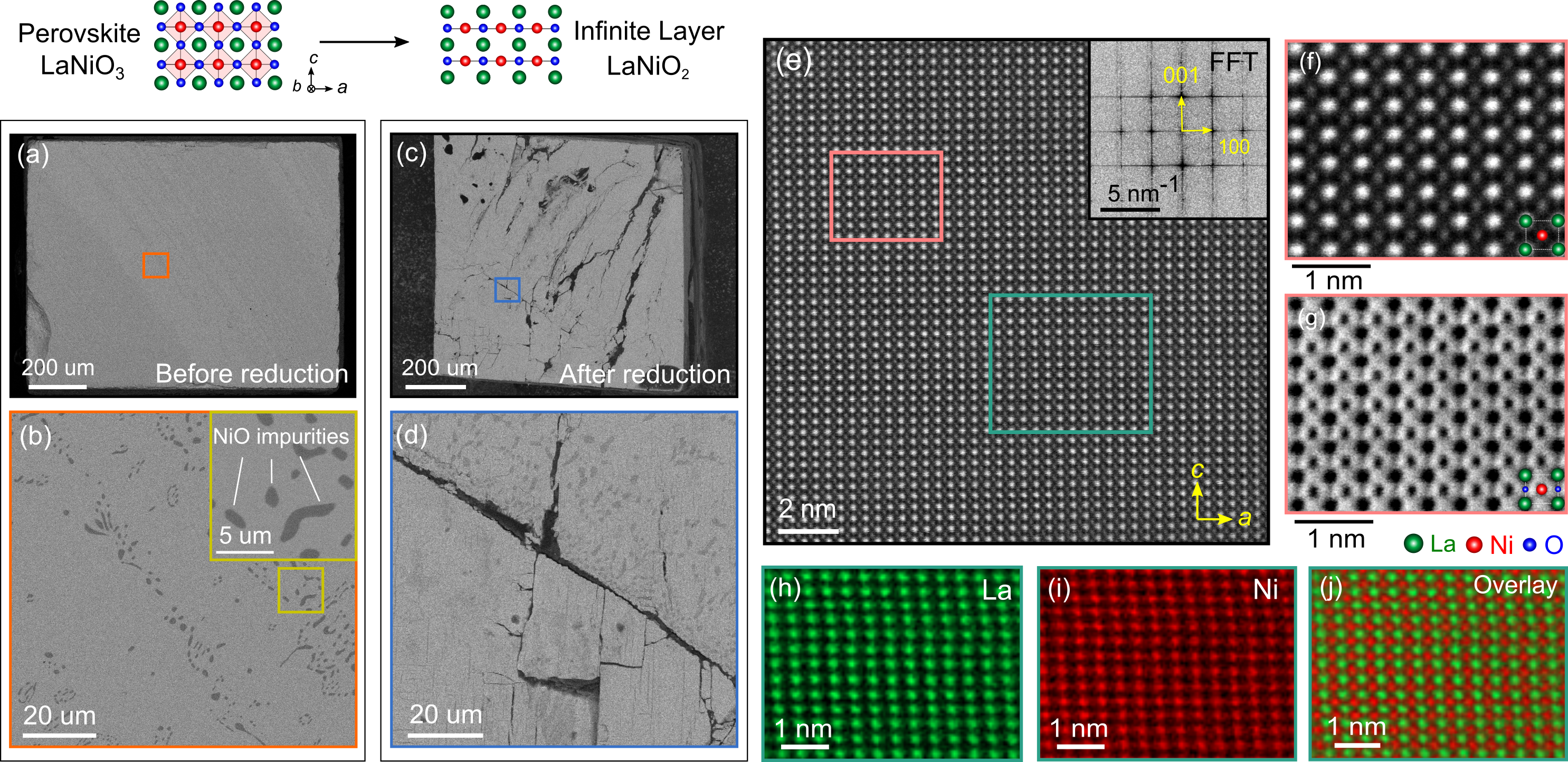}
\caption{\label{Fig1}(a,c) SEM-secondary electron (SE) image of the perovskite LaNiO$_3$ and reduced LaNiO$_2$ crystals, respectively. (b,d) Enlarged views of the images from the orange and blue squares in (a) and (c), respectively. Inset zooms in the region from the yellow square in (b), showing NiO impurities in the crystal. (e) Atomic-resolution STEM-HAADF image taken in the LaNiO$_2$ crystal after topotactic reduction. The inset shows the corresponding fast Fourier transform pattern. (f),(g) The simultaneously acquired STEM-HAADF and ABF images of the region from the red rectangle in (e). (h)-(j) STEM-EELS elemental maps of La, Ni, and the overlaid map, obtained from the green rectangle in (e).}
\end{figure*}

Figures \ref{Fig1}(a)-\ref{Fig1}(d) present scanning electron microscopy (SEM)-backscattered electron (BSE) images of the crystals before and after the topotactic reduction.
A top-down view of a LaNiO$_3$ crystal with typical lateral dimensions of around 1 mm shows a polished surface without any visible domain boundaries [Fig. \ref{Fig1}(a)]. Upon closer inspection within a zoomed-in area, the emergence of an irregularly shaped impurity phase, characterized by a darker contrast owing to the presence of NiO impurities within the crystal, becomes evident  [Fig. 1(b)] \cite{puphal2023phase}.
While a LaNiO$_2$ crystal with a minimal amount of NiO inclusions was selected for the study in Ref.~\cite{hayshida2023}, for the present study, we chose a crystal with a high impurity density.
Figures \ref{Fig1}(c) and \ref{Fig1}(d) show the representative SEM-BSE images of a LaNiO$_2$ single crystal. Discernible impurity phases are observed within various regions of this reduced crystal [Fig. \ref{Fig1}(d)]. Furthermore, several adjacent cracks and dark lines oriented in a parallel or orthogonal direction are present, with a distance and a length ranging from a few to ten micrometers, similar to the pattern previously observed in La$_{1-x}$Ca$_x$NiO$_2$ crystals \cite{puphal2021}. 
These cracks and lines on the crystal surface correspond to the dark amorphous GB-like regions with a width of a few hundred nanometers, partitioning LaNiO$_2$ domains within a cross-sectional TEM specimen (see Fig. S1 of the Supplemental Material for more details). 
The existence of these cracks after reduction indicates a consequence of topotactically induced structural changes.

To understand the local structural changes upon reduction in more detail, we first investigate the atomic lattice in the LaNiO$_2$ crystal by atomic-resolution STEM high-angle annular dark-field (HAADF) imaging. 
Identical TEM specimens from both impurity-free and impurity-containing regions of LaNiO$_2$ were prepared for comparison.
Within a single-crystalline domain in the region without any impurities, the STEM-HAADF image presents a high crystalline quality without any defects or impurity phase after the reduction process [Fig. \ref{Fig1}(e)].
Furthermore, images with a larger field of view do not reveal any regions with defects or impurity phases in one individual domain (Fig. S1). The typical domain size within the crystal is found to be at a scale of a few tenths of nanometers. 
The crystal symmetry is shown by the corresponding fast Fourier transforms (FFT) of the image [Fig. \ref{Fig1}(e) inset]. 
The difference of the distances between the FFT maxima in reciprocal space along the [001] and [100] axis elucidates a tetragonality $c/a$ ratio of approximately 0.8 (highlighted by yellow arrows on FFT patterns). This indicates a tetragonal structure with an out-of-plane lattice contraction and an in-plane lattice expansion upon topotactic removal of apical oxygen atoms. This $c/a$ ratio aligns with the lattice parameters $a,b = 3.9642(3)$~{\AA} and $c = 3.3561(3)$~{\AA} determined by x-ray diffraction for a LaNiO$_{2}$ crystal in Ref.~\cite{hayshida2023}. Further zooming-in STEM-HAADF and annular bright-field (ABF) images are displayed in Fig. \ref{Fig1}(f) and \ref{Fig1}(g). The distribution of oxygen ions including filled and empty apical oxygen sites is clearly visible by ABF imaging. The absence of image contrast at apical oxygen sites confirms the infinite-layer structure of LaNiO$_2$ crystals.
EELS elemental maps of La and Ni obtained from the crystal are shown in Figs. \ref{Fig1}(h)-\ref{Fig1}(j) using La $M_{5,4}$ and Ni $L_{3,2}$ edges, respectively. The maps confirm the uniform distributions of La and Ni throughout the crystal structure. 
This suggests that each domain retains a stoichiometric and high-quality infinite-layer phase within each domain following the reduction process.

%Figure 2
\begin{figure*}[t]
\centering
\includegraphics[width=\textwidth]{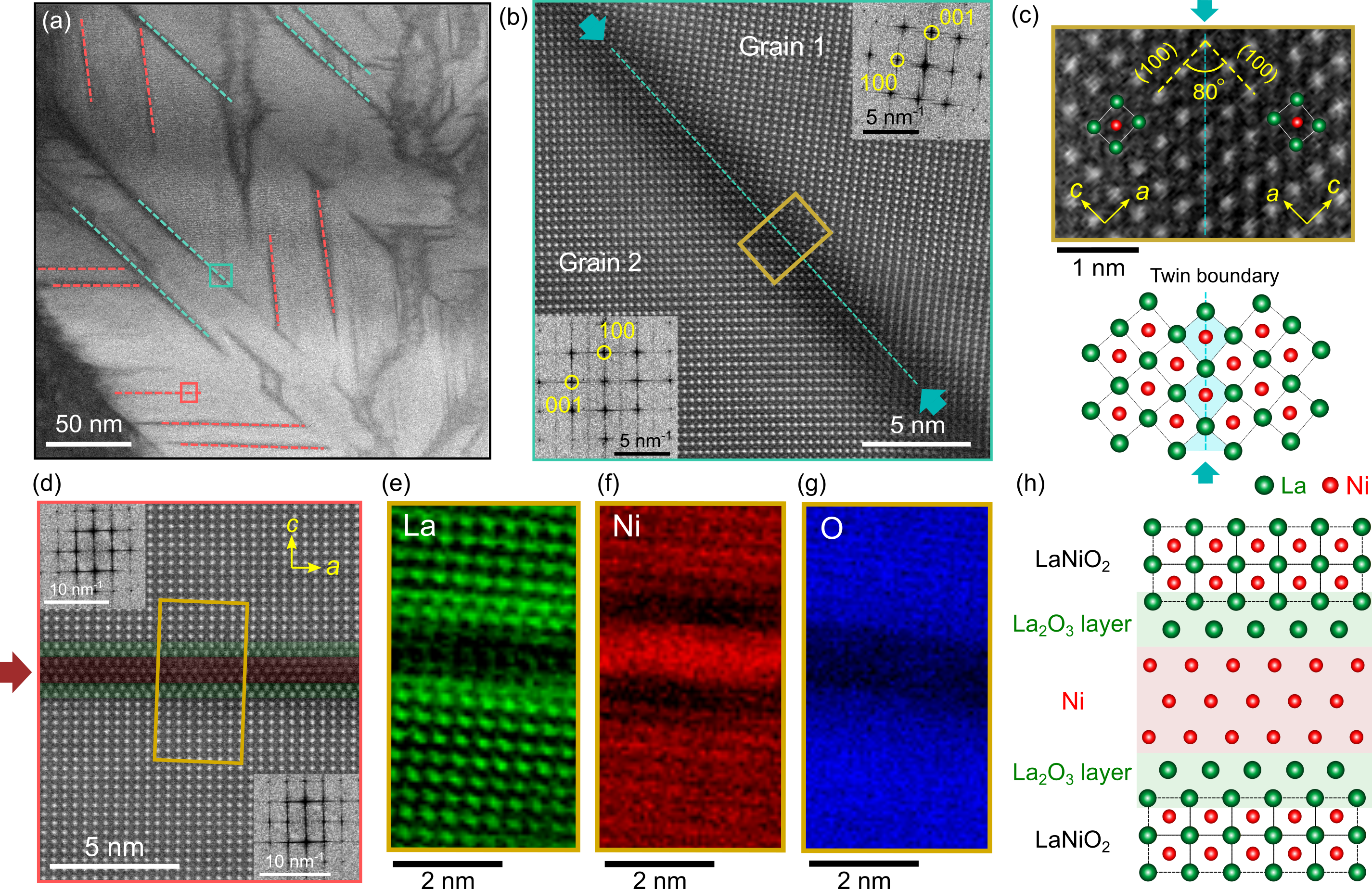}
\caption{\label{Fig2}(a) A low-magnification STEM-HAADF image of a LaNiO$_2$ crystal. The blue and red dashed lines define two types of boundaries observed in the crystal. (b) The zoom-in image of the area from the blue square in (a). The insets are the fast Fourier transform patterns of grains 1 and 2, showing the same structural phases between the boundary. (c) Enlarged view of the area from the yellow rectangle in (b) and a sketch of the structural model for the coherent twin boundary. (d) Enlarged view of the area from the red rectangle in (a). The green and red shaded areas indicate different atomic stacking from the single-crystalline regions. (e)-(g) STEM-EELS elemental maps of La, Ni, O from the yellow rectangle in (d) across the grain boundary. (h) Sketch of the structural model of the grain boundary formed by Ni metal and La$_2$O$_{3-\delta}$ layers.}
\end{figure*}

The alternation in the tetragonality $c/a$ ratio of the crystal through reduction can destabilize the infinite-layer structure in LaNiO$_2$, potentially causing structural inhomogeneities or defects \cite{puphal2023}.
Hence, our focus shifts toward comprehending the structural changes in the crystal after the reduction process.
At a macroscopic scale, the crystal presents a network of GB-like regions arranged orthogonally, suggesting that LaNiO$_2$ retains aligned domains with $a$ or $c$-axis orientations globally through reduction (Fig. S1).
This implies an out-of-plane (in-plane) lattice contraction (expansion) of the crystal upon reduction paralleling any of the three symmetry-related [100] axes of the pseudocubic perovskite phase.
Zooming into a low-magnification STEM-HAADF image of LaNiO$_2$ in Fig. \ref{Fig2}(a), we observe inclined and intersecting dark features.
The dashed lines denote two types of boundaries: lines oriented diagonally (blue); and lines oriented vertically and horizontally (red).
Magnification of an area along one of the dark lines [Fig. \ref{Fig2}(b)], as denoted by the blue dashed line and arrows, reveals two grains oriented along [100] and [001] directions [Fig. \ref{Fig2}(b), insets]. These two crystalline grains, labeled 1 and 2, exhibit an identical lattice structure but with a rotation relative to one another around the misorientation axis. 
In a detailed view of the boundary between two grains in Figure \ref{Fig2}(c), grain 2 appears rotated $\sim$80$^\circ$ around the [001] axis with respect to grain 1. 
Such twinning, where the crystal structure of one part mirrors the matrix, often emerges during a phase transformation within a material to accommodate the strain \cite{christian1995}.
In LaNiO$_2$ crystals, this arises from the perovskite pseudocubic to a tetragonal phase transition induced by topotactic reduction. Using simple geometrical calculations, the rotation angle, $\theta$, is determined as: $\theta = 2 \tan^{-1}(c/a)$. The observed $c/a$ ratio of $\sim$0.8 from the FFT maxima in Fig. \ref{Fig1}(e) aligns with a rotation angle of 80$^\circ$ between the matrix around the [001] axis. 
This is consistent with the twinning angle observed from the image and confirms that the pseudocubic to tetragonal phase transformation resulting in an internal strain, causes twinning throughout the crystal.
The dark contrast of the boundary can be attributed to diffuse and dynamic scattering resulting from the de-channeling of the electron beam as a consequence of the difference in the structure of the boundary compared to the bulk domains \cite{WANG1994,yu2004study}.

%Figure 3
\begin{figure*}[t]
\centering
\includegraphics[width=0.95\textwidth]{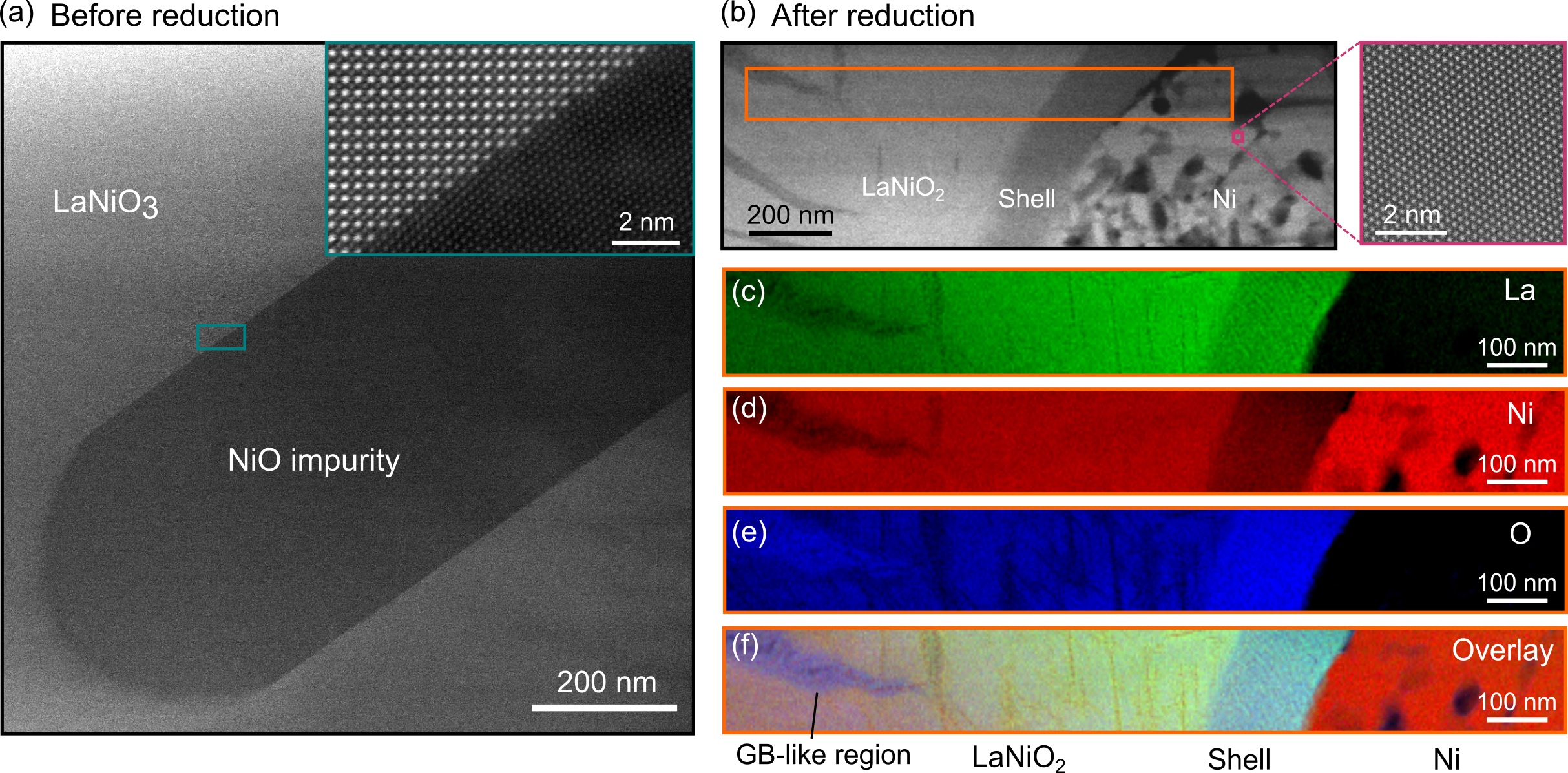}
\caption{\label{Fig3}(a) Low-magnification STEM-HAADF image showing the NiO impurity in the LaNiO$_3$ crystal. The inset shows a magnified image of the interface region between LaNiO$_3$ and the impurity from the green rectangle of the blue rectangle in (a). (b) Low-magnification STEM-HAADF image showing the Ni inclusion surrounded by a shell in the LaNiO$_2$ crystal. The right panel shows the magnified view of the Ni inclusion. (c)-(f)  STEM-EELS elemental maps of La, Ni, O and the superimposed map obtained from the orange rectangle in (b), including crystalline and grain boundary-like regions in LaNiO$_2$, oxygen-rich shell and Ni.}
\end{figure*}

On the other hand, as indicated by the red dashed lines in Fig. \ref{Fig2}(a), a different type of boundary becomes evident, aligned parallel to crystallographic directions. As shown in Fig. \ref{Fig2}(d), the dark line aligns along the [100] direction of LaNiO$_2$.  
Contrary to the grains connected through a twin boundary as shown in Fig. \ref{Fig2}(b)-\ref{Fig2}(c), grains here share an interfacial transition region structurally and chemically distinct from their adjacent counterparts.
Specifically, a Ni layer with a cubic rock-salt structure is sandwiched between additional La-O layers exhibiting a horizontal relative shift by $a$/2[100].
At the GB region, the spacing between the La-O layer and its adjacent one ($\sim$2.8 \AA) is smaller than the spacing between two neighboring La-O layers ($\sim$3.4 \AA) in the bulk LaNiO$_2$ phase. 
To confirm the chemical composition, STEM-EELS elemental maps of La, Ni, and O across the GB are displayed in Fig. \ref{Fig2}(e)-\ref{Fig2}(g). The La map reveals the extra LaO$_{1-\delta}$ layers segregating the LaNiO$_2$ from the inner Ni layer. The region with enhanced Ni content indicates the presence of the Ni layer at the GB, while the oxygen intensity decreases at the GB region.
Previous powder x-ray diffraction measurements Ref. \cite{puphal2023} have identified a chemical decomposition of LaNiO$_2$ crystals into elemental Ni, La$_2$O$_3$, and LaOH due to excessive reduction. Consequently, our findings likely reveal the decomposition process associated with the formation of GB containing elemental Ni bordered by La$_2$O$_{3-\delta}$ blocks between domains.

Having discussed the local structural deformations, we now turn to the larger-scale imperfections in the crystal due to topotactic reduction. As shown in Fig. \ref{Fig1}(b) and \ref{Fig1}(d), impurity phases are present in both unreduced and reduced crystals. To discern the origin of these impurities, we first investigate the unreduced crystal.  
The low-magnification STEM-HAADF image of LaNiO$_3$ in Fig. \ref{Fig3}(a) reveals a pristine crystalline region housing a rod-shaped NiO impurity. 
Upon closer inspection, the atomic structure of cubic rock-salt NiO is in close contact with pseudocubic perovskite LaNiO$_3$ with a structurally coherent interface due to closely aligned lattice symmetries [Fig. \ref{Fig3}(a), inset]. Supporting STEM-HAADF and EELS maps in Fig. S2 confirm the NiO impurity phase and a chemically sharp LaNiO$_3$/NiO interface.
After reduction, an extent of decomposition near the impurity and within the crystalline region throughout the crystal appears, while a partially polycrystalline character is present inside the Ni inclusion [Fig. \ref{Fig3}(b)]. 
Zooming in on a single grain exhibits a cubic rock-salt structure within the impurity phase. 
Interestingly, a dark shell with a width of approximately a hundred nanometer surrounding the Ni inclusion separates the impurity phase from the crystalline region, unlike the sharp LaNiO$_3$/NiO interface observed pre-reduction.
Within the LaNiO$_2$ bulk region, some vertical lines and irregular dark GB-like regions with a width of a few ten nanometers are observed. Both the shell and GB-like region exhibit an amorphous structure, displaying dark contrast in the image owing to the diffuse scattering \cite{WANG1994}.

In Fig. \ref{Fig3}(c)-\ref{Fig3}(f), the elemental maps reveal how the crystal structure is modified due to the reduction process: a Ni-rich phase emerges without other contributions confirming that the NiO impurity in LaNiO$_3$ is reduced to Ni metal in LaNiO$_2$. 
An excess of Ni accompanying La and O deficiencies are also present at the vertical lines in the bulk region, indicating a formation of Ni metal layers that are identical to the structural modifications observed upon excess reduction (c.f., Fig. \ref{Fig2}(d)). Besides, the O elemental map in Fig. \ref{Fig3}(e) provides a clear view of enhanced intensity in the amorphous shell and GB-like regions.
These findings suggest that the removal of oxygen drives the transformation of NiO impurities into polycrystalline Ni inclusions. 
The observed structural disorder within the surrounding shell and GB-like region blocks oxygen to be pulled out in the crystal during the reduction and, therefore appears to be oxygen-rich relative to stoichiometric LaNiO$_2$ domains, leading to the formation of an amorphous LaNiO$_{2+\delta}$ phase (Fig. S3).
Such a disordered phase can potentially be associated with a mechanism relieving reduction-induced strain as well. Functioning as a transition layer, the LaNiO$_{2+\delta}$ phase between the domains and Ni impurities and in the crystal relax the local strain arising from the lattice mismatch between the defect and bulk regions.

\section{Conclusion}
In summary, we scrutinize how the topotactic oxygen reduction impacts the structural transformation of LaNiO$_3$ to the infinite-layer phase. 
Formation of macroscopic and local structural imperfections are observed in response to the reduction-induced crystal symmetry changes in LaNiO$_2$. 
At the nanometer scale, we discern two types of boundaries separating domains. Globally, the infinite-layer phase is preserved in LaNiO$_2$, featuring coherent twin boundaries oriented diagonally to accommodate the internal strain.
Besides, non-stoichiometric grain boundaries coexist within the crystal, alongside a decomposition into La$_2$O$_{3-\delta}$ and Ni metal due to excessive reduction, formed in parallel to crystallographic orientations between crystalline regions. 
Moving to the micrometer scale, reduction-induced structural disorder surrounds the Ni metal impurities and resides within crystalline domains. These disordered regions, impeding the local oxygen reduction, result in a transition LaNiO$_{2+\delta}$ phase. 
Our atomic-scale observation of these defects provides a good basis for comprehending the reduction mechanism and offers valuable insights into optimizing the synthesis of nickelate single crystals and related materials in future endeavors.

\begin{acknowledgments}
We thank J. Deuschle for TEM specimen preparation. This project has received funding in part from the European Union’s Horizon 2020 research and innovation program under Grant Agreement No. 823717–ESTEEM3.
\end{acknowledgments}

\nocite{*}
%\bibliography{Ref}% Produces the bibliography via BibTeX.

%

\end{document}